\documentclass[aip,jcp,reprint,amsmath,amssymb]{revtex4-1}

\usepackage{graphicx} 
\usepackage{dcolumn}
\usepackage{bm}
\usepackage[mathlines]{lineno}

\usepackage[utf8]{inputenc}
\usepackage[T1]{fontenc}
\usepackage{mathptmx}
\usepackage{etoolbox}
\usepackage{hyperref}
\hypersetup{
    hidelinks
}

\draft

\makeatletter
\def\@email#1#2{%
 \endgroup
 \patchcmd{\titleblock@produce}
  {\frontmatter@RRAPformat}
  {\frontmatter@RRAPformat{\produce@RRAP{*#1\href{mailto:#2}{#2}}}\frontmatter@RRAPformat}
  {}{}
}%
\makeatother

\begin{document}

\title{Phase-Field Models for Particle-Stabilised Emulsions}

\author{Elisabeth C. Eij }
\affiliation{Van 't Hoff Laboratory for Physical and Colloid Chemistry, Utrecht University, Utrecht, The Netherlands}
\affiliation{Institute for Theoretical Physics, Utrecht University, Utrecht, The Netherlands}

\author{Joost de Graaf}
\affiliation{Institute for Theoretical Physics, Utrecht University, Utrecht, The Netherlands}

\author{Martin F. Haase*}
\email[]{m.f.haase@uu.nl}
\affiliation{Van 't Hoff Laboratory for Physical and Colloid Chemistry, Utrecht University, Utrecht, The Netherlands}

\author{Jesse M. Steenhoff**}
\email[*]{j.m.steenhoff@uu.nl}
\affiliation{Van 't Hoff Laboratory for Physical and Colloid Chemistry, Utrecht University, Utrecht, The Netherlands}

\date{18 February 2026}

\begin{abstract}
Particle-stabilised emulsions are a cornerstone of soft matter science due to their broad application and fundamental relevance. Computer simulations provide key insights into the formation and behaviour of these emulsions, yet current methods are limited by the spatiotemporal scales accessible for study. The principal issue is that particles are resolved individually. In this work, an alternative strategy is introduced based on phase-field theory, for which we establish the framework. By evolving continuous fields, large-scale dynamics can be simulated in a computationally efficient manner. Our approach is then applied to model the complex formation of a bicontinuous interfacially jammed emulsion gel (bijel) \textit{via} solvent-transfer induced phase separation (STrIPS). By resolving the coupled dynamics of liquid phase separation and nanoparticle adsorption, the model allows for the characterisation of the influence of nanoparticles on the morphology. Higher concentrations of nanoparticles are found to reduce the average domain size of STrIPS bijels, in line with previous experimental evidence. The presented phase-field model thus represents a promising approach for the morphological investigation of complex particle-stabilised emulsions. 

\end{abstract}

\maketitle

\section{Introduction}
\begin{figure}
    \centering
    \includegraphics[width=\linewidth]{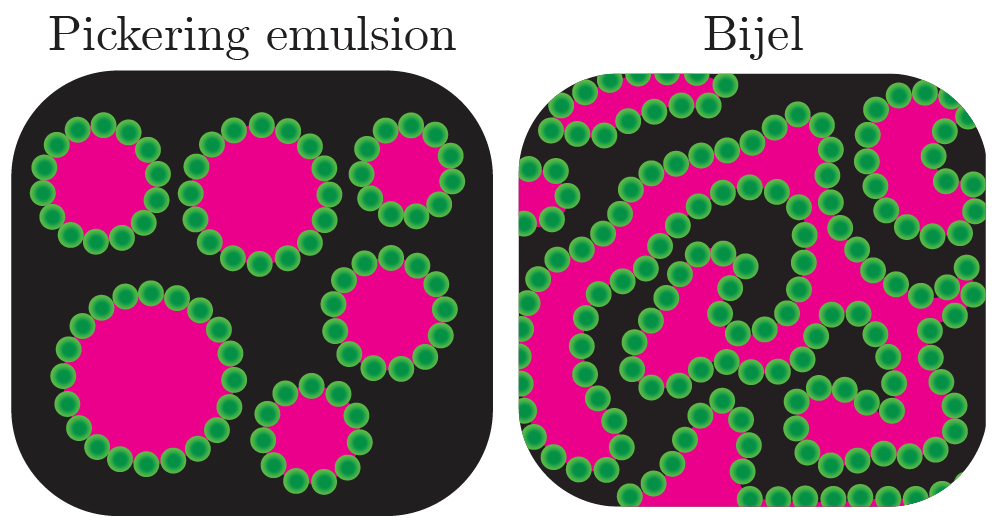}
    \caption{Schematic 2D representations of the different morphologies of particle-stabilised emulsions. The Pickering emulsion (left) is made up of dispersed droplets in a continuous medium, with nanoparticles occupying the surface of the droplets. In contrast, the bijel (right) is a continuous structure of interwoven liquid channels, separated by a percolating sheet of nanoparticles.}
    \label{PickeringAndBijel}
\end{figure}

Similar to surfactants, nanoparticles can attach to the interface formed by the phase separation of immiscible liquids\cite{Ramsden1904,Pickering1907,Binks2003}. By significantly reducing further phase separation through coalescence\cite{Wiley1954,Tambe1994,Arditty2003,Pawar2011} and Ostwald ripening \cite{Ashby2000,Meinders2004,Cervantes2008,Tcholakova2008,Binks2010}, these nanoparticles create emulsions that display greatly enhanced stability relative to their surfactant-stabilised counterparts. This robustness makes such particle-stabilised emulsions of particular interest for applications, further supplemented by the option of tailoring the specific application to a suitable emulsion morphology. In general, the morphologies of the particle-stabilised emulsions can be categorised into two classes, which are exemplified in Figure~\ref{PickeringAndBijel}. 

The more common morphology, depicted on the left in Figure~\ref{PickeringAndBijel}, consists of particle-decorated droplets dispersed in a continuous liquid medium. This type of particle-stabilised emulsion, generally known as a Pickering emulsion \cite{Pickering1907}, lends itself particularly well to the purpose of encapsulation and delivery, finding ample application in coatings\cite{Haase2012}, cosmetics\cite{Wu2020,Wei2020} and pharmaceutics\cite{Simovic2007,Frelichowska2009}. Alternatively, nanoparticles can kinetically stabilise the bicontinuous morphology formed during the spinodal decomposition of immiscible liquids. The resulting material, shown on the right in Figure~\ref{PickeringAndBijel}, is known as a bicontinuous interfacially jammed emulsion gel (bijel)\cite{Stratford2005,Herzig2007}. Rather than the compartmental nature of the Pickering emulsion, applications of the bijel exploit the continuity found in its interwoven network of liquid channels, with examples in catalysis\cite{Cha2019,Vitantionio2019,Park2021}, energy storage\cite{Witt2016,Cai2018,Garcia2019,Gross2021} and membrane separation\cite{Haase2017,Siegel2022}. 

Although different in appearance and target function, both the Pickering emulsion and the bijel share a common origin as particle-stabilised emulsions\cite{Binks2003}. Their overall structure, as well as their behaviour, is dictated by their formative mechanism of phase separation and nanoparticle attachment. Considering its relevance to the effective application of particle-stabilised emulsions, extensive investigation into this interplay between the phase separation of immiscible liquids and the interfacial attachment of nanoparticles is warranted. Computer simulations, with a proven track record for studying particles on liquid interfaces\cite{Bresme2007,Harting2014,Morris2015,Dasgupta2017,Thijssen2018}, are especially suitable for this purpose. 

In fact, both the Pickering emulsion and bijel have an expansive history with computer simulations. The bijel in particular, first identified in simulations\cite{Stratford2005} two years prior to its experimental realisation\cite{Herzig2007}. Previous works on these systems generally deal with either the micro- or mesoscale dynamics of nanoparticles at liquid interfaces, employing simulation techniques relevant to the respective spatiotemporal scales. Common examples include (coarse-grained) molecular dynamics\cite{Ranatunga2011,Geng2023}, dissipative particle dynamics\cite{Hore2007,Fan2012,Zhao2016} or Lattice-Boltzmann\cite{Stratford2005,Kim2008,Joshi2009,Jansen2011,Frijters2014,Schiller2024}. Such techniques are very powerful, facilitating studies into the morphological effect of particle properties such as volume fraction\cite{Frijters2014}, shape\cite{Schiller2024} and wettability\cite{Jansen2011,Fan2012}. The inclusion of individual particles, however, puts a limit on the spatiotemporal scales accessible by the simulations. In the mesoscale case with micrometer-sized particles, these scales typically only go up to micrometers/milliseconds\cite{Schiller2024}. For nanoparticles these restrictions are even more severe, with the length- and timescales being sub-micron and sub-microsecond, respectively\cite{Jansen2011}. For particle-stabilised emulsions, however, there are instances where the scales of interest are considerably larger. A prime example of such an instance is the formation of bijels \textit{via} solvent-transfer induced phase separation (STrIPS). 

In STrIPS\cite{Haase2015}, the diffusion of a solvent triggers the phase separation of an immiscible oil/water mixture \textit{via} spinodal decomposition. Nanoparticles subsequently attach to the resulting liquid-liquid interface, eventually arresting further phase separation and stabilising the bijel structure. During this process, the bijel evolves characteristic morphological features on the spatiotemporal scales associated with the macroscopic diffusion of the solvent. For example, bijels fabricated \textit{via} STrIPS display a prominent gradient in the size of their liquid domains, being formed over the course of seconds and extending over hundreds of micrometers\cite{Khan2022,Siegel2024}. On these large scales a typical STrIPS system already involves billions of nanoparticles. Consequently, resolving individual particle interactions \textit{via} mesoscopic techniques is computationally unattainable. This problem could, however, be addressed through alternative means: implementing a phase-field method. 

Phase-field modelling represents key physical processes through coupled sets of partial differential equations. Phase-field models have been widely used for systems with mass transfer dynamics similar to those of STrIPS, such as polymeric membrane formation through phase inversion\cite{Zhou2006,Cervellere2019,Cervellere2021}, but the full extension to particle-stabilised emulsions has not yet been realised. Although there have been several attempts in literature to apply phase-field models to particle-stabilised emulsions\cite{Millet2011, Aland2011, Aland2012, Carmack2015, Carmack2017, Carmack2018}, these generally still resolve individual particles, albeit at a phase-field level. 

The inclusion of individual particles prevents the effective scale-up to systems with a size relevant for the STrIPS process. More recent work\cite{Steenhoff2025} of ours overcomes this problem by abandoning the inclusion of nanoparticles altogether, instead using the interfacial tension as a substitute measure for arresting phase separation of the bulk liquids. A more accurate description of the STrIPS process, however, should do more justice to its interfacial aspects and prominently feature the influence of nanoparticles on the resulting morphology.

In this work, this is achieved by introducing a phase-field formalism for the formation of particle-stabilised emulsions. First, sets of partial differential equations are derived that govern the interfacial accumulation and jamming of nanoparticles during liquid-liquid phase separation. By treating the nanoparticles as a continuous field, rather than discrete objects, their presence can be taken into account while remaining computationally efficient. Subsequently, it is shown that the attached nanoparticles effectively prevent further phase separation, suppressing Ostwald ripening. Finally, the phase-field model is applied to the formation of bijels \textit{via} STrIPS, revealing a strong morphological influence of the nanoparticle concentration that matches previous experimental findings. 

\section{Theory}
\subsection{Free-Energy Functional \& Chemical Potentials}
Based on the phase-field theory originally developed for surfactants\cite{Toth2015}, the thermodynamics of an inhomogeneous mixture of immiscible liquids ($\phi$) containing nanoparticles ($\psi$) are contained in the following free-energy functional

\begin{equation}
    F=\int_V dV \left\{\mathcal{F}_\phi(\phi)+\mathcal{F}_\psi(\psi)+\mathcal{F}_I(\nabla\phi,\psi) \right\}.
    \label{FreeEnergyFunctional}
\end{equation}

Herein, three main components of the system free energy are distinguished. Owing to the large size discrepancy between the molecular liquid and the nanoparticles, their bulk contributions to the free-energy density, respectively $\mathcal{F}_\phi(\phi)$ and $\mathcal{F}_\psi(\psi)$, are considered independent. For the former, a Flory-Huggins expression captures the behaviour of a binary liquid undergoing phase separation
\begin{equation}
    \mathcal{F}_\phi(\phi)=f\left(\phi\ln\phi+(1-\phi)\ln(1-\phi)+\chi\phi(1-\phi)\right),
    \label{FloryHuggins}
\end{equation}
where $f=k_BT/V$ is the characteristic scale of the free-energy density, with the Boltzmann constant $k_B$, the absolute temperature $T$ and the system volume $V$. The degree of interaction between the immiscible liquids is determined by the parameter $\chi$, with phase separation initiated for $\chi>2$. For the strongly immiscible liquids considered in this work, an interaction parameter of $\chi=3$ is an appropriate choice\cite{Zhang2021}. 

Neglecting interparticle interactions, an ideal mixture approximation gives the bulk free-energy density of the nanoparticles as
\begin{equation}
    \mathcal{F}_\psi(\psi)=f\psi(\ln\psi-1).
\end{equation}

Lastly, $\mathcal{F}_I(\nabla\phi,\psi)$ is the component of the free-energy density associated with the presence of liquid interfaces in the system. Containing contributions from both the pure liquids and the nanoparticles, this component takes the shape of 

\begin{equation}
    \mathcal{F}_I(\nabla\phi,\psi)=\frac{\kappa}{2}\lvert\nabla\phi\rvert^2-\frac{\alpha}{2}\psi\lvert\nabla\phi\rvert^2.
\end{equation}
Here, $\frac{\kappa}{2}\lvert\nabla\phi\rvert^2$ is the gradient energy from Cahn-Hilliard theory\cite{Cahn1958},  energetically penalising liquid interfaces through the gradient coefficient $\kappa$. This gradient coefficient $\kappa$ also introduces a characteristic length to the system, namely the interfacial width $\lambda=\sqrt{\kappa/f}$. In contrast to the first term, the second term $-\frac{\alpha}{2}\psi\lvert\nabla\phi\rvert^2$ reduces the free energy by coupling liquid interfaces and nanoparticles through the attachment parameter $\alpha$. This latter term effectively renders the nanoparticles surface-active, prompting interfacial accumulation for $\alpha>0$.

With the free-energy functional established, the chemical potentials of the liquid and nanoparticles directly follow from its variational derivatives. For the liquid, the chemical potential is thus given by 
\begin{equation}
    \mu_{\phi}=\frac{\delta F}{\delta\phi}=\frac{\partial \mathcal{F}_\phi}{\partial\phi}-\nabla\cdot\frac{\partial \mathcal{F}_I}{\partial(\nabla\phi)};
\end{equation}
\begin{equation}
    \mu_\phi=\frac{\partial \mathcal{F}_\phi}{\partial\phi}-(\kappa-\alpha\psi)\nabla^2\phi+\alpha\left(\nabla\phi\cdot\nabla\psi\right),
\end{equation}  
with the partial derivative of the Flory-Huggins bulk contribution as 
\begin{equation}
    \frac{\partial \mathcal{F}_\phi}{\partial\phi}=f\left(\ln{\frac{\phi}{1-\phi}} +\chi(1-2\phi)\right).
\end{equation}
Analogously, the chemical potential of the nanoparticles becomes 
\begin{equation}
    \mu_{\psi}=\frac{\delta F}{\delta\psi}=\frac{\partial \mathcal{F}_\psi}{\partial\psi}+\frac{\partial \mathcal{F}_I}{\partial\psi};
\end{equation}
\begin{equation}
    \mu_\psi=f\ln\psi-\frac{\alpha}{2}\lvert\nabla\phi\rvert^2.
\end{equation}

\subsection{Multi-Component Cahn-Hilliard Equations}
Neglecting convection in the system, a simplification addressed in the Discussion section, mass transport solely occurs through diffusion. Accordingly, the system evolves through diffusional fluxes driven by gradients in the chemical potential. By combining these diffusional fluxes with a continuity expression for mass conservation, the dynamics of the resulting system are described by the multi-component Cahn-Hilliard equation 

\begin{equation}
    \frac{\partial\phi_i}{\partial t}=\nabla\cdot\sum_jM_{ij}\nabla\mu_j\ ,
\end{equation}
where $M_{ij}$ is the mobility of component $i$ due to a gradient in the chemical potential of component $j$. Considering the size difference between the nanoparticles and the molecular liquids, the cross-terms where $i\neq j$ can be safely ignored. The governing equations of the liquids and the nanoparticles thus become 
\begin{equation}
    \frac{\partial\phi}{\partial t}=\nabla\cdot(M_\phi\nabla\mu_\phi);
    \label{CahnHilliard1}
\end{equation}
\begin{equation}
    \frac{\partial\psi}{\partial t}=\nabla\cdot(M_\psi\nabla\mu_\psi).
    \label{CahnHilliard2}
\end{equation}
Under the additional assumption that the mobilities of the liquid and the nanoparticles, respectively $M_\phi$ and $M_\psi$, are constants across space, these equations reduce to 
\begin{equation}
    \frac{\partial\phi}{\partial t}=M_\phi\nabla^2\mu_\phi;
    \label{CahnHilliard3}
\end{equation}
\begin{equation}
    \frac{\partial\psi}{\partial t}=M_\psi\nabla^2\mu_\psi.
    \label{CahnHilliard4}
\end{equation}

Prior to their numerical solution, Eqs.~(\ref{CahnHilliard3}) and (\ref{CahnHilliard4}) are rendered dimensionless through the following variables
\begin{equation}
    \tilde{\textbf{x}}=\frac{\textbf{x}}{\lambda}=\textbf{x}\sqrt{\frac{f}{\kappa}};
\end{equation}
\begin{equation}
    \tilde{t}=\frac{t}{\tau}=t\frac{f^2M_\phi}{\kappa}.
\end{equation}
Given that it naturally emerges from the formulation of the free energy, the interfacial width $\lambda=\sqrt{\kappa/f}$ is taken as the characteristic length scale of the system. The liquids display the fastest dynamics of interest, so time is scaled with respect to their characteristic diffusion time $\tau= \kappa/(f^2M_\phi)$  over the interfacial length $\lambda$.

In dimensionless form, the governing equations for the liquids and the nanoparticles become 
\begin{equation}
    \frac{\partial\phi}{\partial \tilde{t}}=\tilde\nabla^2\tilde{\mu}_\phi;
    \label{ScaledCH1}
\end{equation}
\begin{equation}
    \frac{\partial\psi}{\partial \tilde{t}}=\tilde{M}_\psi\tilde\nabla^2\tilde\mu_\psi\ ,
    \label{ScaledCH2}
\end{equation}
with $\tilde M_\psi=M_\psi/M_\phi$ as the relative mobility of the nanoparticles with respect to the liquid. Additionally, $\tilde\mu_\phi$ and $\tilde \mu_\psi$ are the dimensionless versions of the chemical potentials 
\begin{equation}
    \tilde{\mu}_\phi=\ln{\frac{\phi}{1-\phi}} +\chi(1-2\phi)-(1-\tilde\alpha\psi)\tilde\nabla^2\phi+\tilde\alpha(\tilde\nabla\phi\cdot\tilde\nabla\psi);
\end{equation}
\begin{equation}
    \tilde{\mu}_\psi=\ln\psi-\frac{\tilde\alpha}{2}\lvert\tilde\nabla\phi\rvert^2,
\end{equation}
where the parameter $\tilde\alpha=\alpha/\kappa$ determines the relative change in energy for the attachment of nanoparticles compared to the formation of a liquid interface.

The dimensionless Eqs.~(\ref{ScaledCH1}) and (\ref{ScaledCH2}) can subsequently be solved numerically \textit{via} a simple finite-volume scheme, the details of which can be found in the ESI. However, while these equations do capture the desired dynamics, their current form also severely restricts their functionality. In order for them to be applicable to particle-stabilised emulsions, these constraints have to be resolved first.

\subsection{Limitations of the Cahn-Hilliard Equations}
\begin{figure*}
    \centering
    \includegraphics[width=\linewidth]{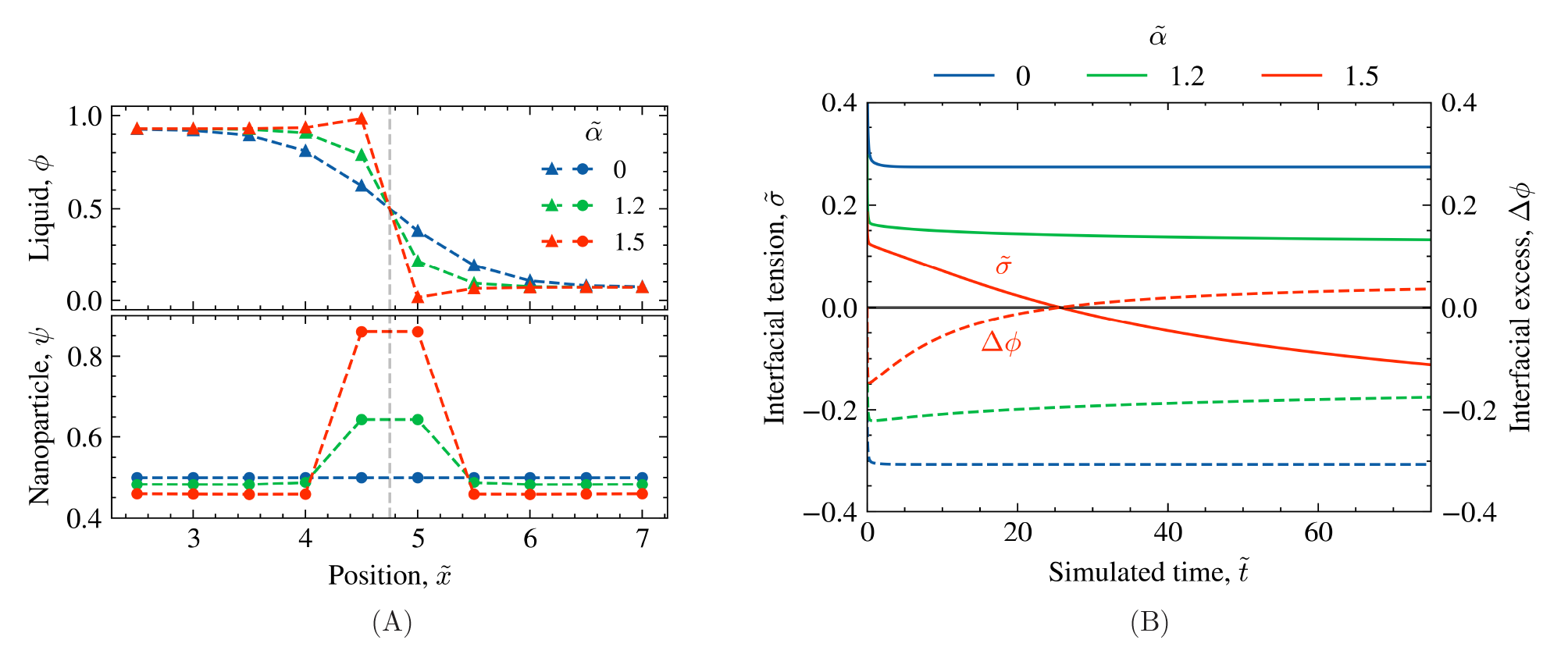}
    \caption{(A) Equilibrium solutions of governing Eqs.~(\ref{ScaledCH1}) and (\ref{ScaledCH2}) for a 1D liquid interface, using different values of the attachment parameter $\tilde{\alpha}$. In both panels, the dashed grey line indicates the centre of the interface. The top and bottom panels show the density profiles of the liquid $\phi$ and nanoparticles $\psi$, respectively. For the sake of illustrating the interface, space is discretised with $\Delta\tilde{x}=0.5$ rather than the usual $\Delta\tilde{x}=1$. (B) Evolution of the interfacial tension (solid lines) and interfacial excess (dashed lines) while forming the interfaces in (A). The interfacial tension is calculated \textit{via} Eq.~(\ref{IFT}), while the interfacial excess here is defined as the difference in liquid composition between the left bulk phase and the interface at $\tilde x=4.5$.}
    \label{CH_Profiles&IFT}
\end{figure*}
The limitations of Eqs.~(\ref{ScaledCH1}) and (\ref{ScaledCH2}) become apparent in their equilibrium solutions for a 1D liquid interface. Figure~\ref{CH_Profiles&IFT}A shows the corresponding density profiles of the liquid $\phi$ and the nanoparticles $\psi$, for increasing values of the attachment parameter $\tilde\alpha$. The top panel illustrates the equilibrium shape of the liquid interface, connecting the compositions of the bulk phases on the left and right. The dashed vertical line represents the centre of the interface, identified by a liquid composition of $\phi=0.50$. The bottom panel contains the density profiles of the nanoparticles, revealing their relative distribution over the bulk and interface of the liquid. 

Whereas for $\tilde{\alpha}=0$ the nanoparticles are homogeneously distributed over the system, they exhibit a marked preference for the interface when $\tilde{\alpha}>0$. This is demonstrated by the higher interfacial density of the nanoparticles $\psi$ relative to the bulk, which becomes more pronounced for higher values of $\tilde{\alpha}$. However, this interfacial accumulation of the nanoparticles simultaneously distorts the equilibrium shape of the liquid interface. As shown in the top panel of Figure~\ref{CH_Profiles&IFT}A, for $\tilde{\alpha}=0$ the liquid interface represents the gradual, monotonic transition between the compositions of the bulk phases. Increasing the local nanoparticle density $\psi$ considerably deforms this interface, becoming sharper and eventually even non-monotonic for higher $\tilde{\alpha}$. 
 
The origin of this behaviour can be traced to the chemical potential of the liquid, in particular the term $(1-\tilde\alpha\psi)\tilde\nabla^2\phi$. This term introduces an ``effective'' version of the gradient energy parameter $\tilde\kappa_{\mathrm{eff}}=1-\tilde\alpha\psi$, which decreases as nanoparticles attach to the liquid interface. As the shape of the liquid interface is partially determined by the gradient energy parameter, a lower value of $\tilde\kappa_{\mathrm{eff}}$ explains the observed increase in its sharpness. 

While the increased sharpness poses a numerical problem if the interface cannot be spatially resolved anymore, there is an additional effect at higher values of $\tilde\alpha$ that proves to be even more detrimental. This effect is directly tied to the interfacial tension of the system, which for a 1D interface with an ``effective'' gradient energy parameter $\tilde\kappa_{\mathrm{eff}}=1-\tilde\alpha\psi$, can be calculated \textit{via}

\begin{equation}
    \tilde\sigma=\int_{-\infty}^{\infty}d\tilde x \left\{(1-\tilde\alpha\psi)\left(\frac{\partial\phi}{\partial \tilde x}\right)^2 \right\}.
    \label{IFT}
\end{equation}

Eq.~(\ref{IFT}) can then be used to track the interfacial tension over the course of the simulations. Using different values of $\tilde\alpha$, the resulting profiles are shown as solid lines in Figure~\ref{CH_Profiles&IFT}B. Figure~\ref{CH_Profiles&IFT}B additionally contains dashed lines that represent the interfacial excess $\Delta\phi$. Defined as $\Delta\phi=\phi(\tilde x=4.5) -\phi_{L}$, it reflects the difference in liquid composition between the interface at $\tilde x=4.5$ and the left bulk phase, with the bulk composition $\phi_L$  determined from Eq.~(\ref{FloryHuggins}). 

For the two lower values of $\tilde\alpha$, the interfacial tension reaches a constant positive value after the equilibrium shape of the interface has been established. Consistent with the difference in the liquid composition over the interface, this is accompanied by a constant negative value of the interfacial excess. For $\tilde\alpha=1.5$, however, the interfacial tension steadily decreases until it eventually becomes negative. At this point, the system can start lowering its free energy by creating more interface, which it achieves here through a positive interfacial excess $\Delta\phi$. Apart from the unphysical nature of the underlying negative interfacial tension, a positive interfacial excess is associated with considerable numerical issues. Not only does it pose a problem for the logarithmic terms in the free-energy functional, it also implies that the nanoparticles simultaneously cause and are attracted to a higher interfacial gradient. The result is a positive feedback loop of nanoparticle attachment that quickly escalates into numerical instability. 

In summary, the form of the Cahn-Hilliard equations proposed in (\ref{ScaledCH1}) and (\ref{ScaledCH2}) induces strong deformation of the liquid interface upon significant accumulation of nanoparticles, rendering them inherently unsuitable for particle-stabilised emulsions. Consequently, a new set of dynamic equations is introduced that can capture the same system dynamics without suffering from interfacial distortions. 

\subsection{Dynamic Equations for Particle-Stabilised Emulsions}
From the discussion in the previous section, it appears that the root cause of the problems with the Cahn-Hilliard approach is that the nanoparticles directly influence the chemical potential of the liquid. This issue can be resolved by partially decoupling the chemical potentials in the system, employing a scaling argument that the chemical potential of the molecular liquid should not be changed by the presence of much larger nanoparticles. The dynamic equations of the liquid and the nanoparticles then respectively become 

\begin{equation}
    \frac{\partial\phi}{\partial \tilde{t}}=\tilde\nabla^2\left(\ln{\frac{\phi}{1-\phi}} +\chi(1-2\phi)-\tilde\nabla^2\phi\right);
    \label{DynamicNoJam1}
\end{equation}

\begin{equation}
    \frac{\partial\psi}{\partial \tilde{t}}=\tilde{M}_\psi\tilde\nabla^2\left(\ln\psi-\frac{\tilde\alpha}{2}\lvert\tilde\nabla\phi\rvert^2\right),
    \label{DynamicNoJam2}
\end{equation}
where the conventional form of the Cahn-Hilliard equation can be recognised for the liquids. Note that the term containing $\lvert\tilde\nabla\phi\rvert^2$ in Eq.~(\ref{DynamicNoJam2}) renders the dynamic equations ``non-integrable'', meaning that this specific combination of equations cannot be formally derived from a free-energy. However, similar simplifications have previously been used to maintain meaningful representations of interfaces in other systems\cite{Mokbel2018}; associated errors can generally be considered negligible for practical purposes. In addition, this method is reminiscent of the so-called Active B model for active-particle phase separation\cite{Wittkowski2014}, albeit with non-integrability being imposed on a set of equations rather than a single one. 

Effectively, the governing Eqs.~(\ref{DynamicNoJam1}) and (\ref{DynamicNoJam2}) describe a system in which nanoparticles simply attach to a scaffold formed by the phase-separating liquids, without actually influencing the scaffold itself.
\begin{figure}
    \centering
    \includegraphics[width=0.98\linewidth]{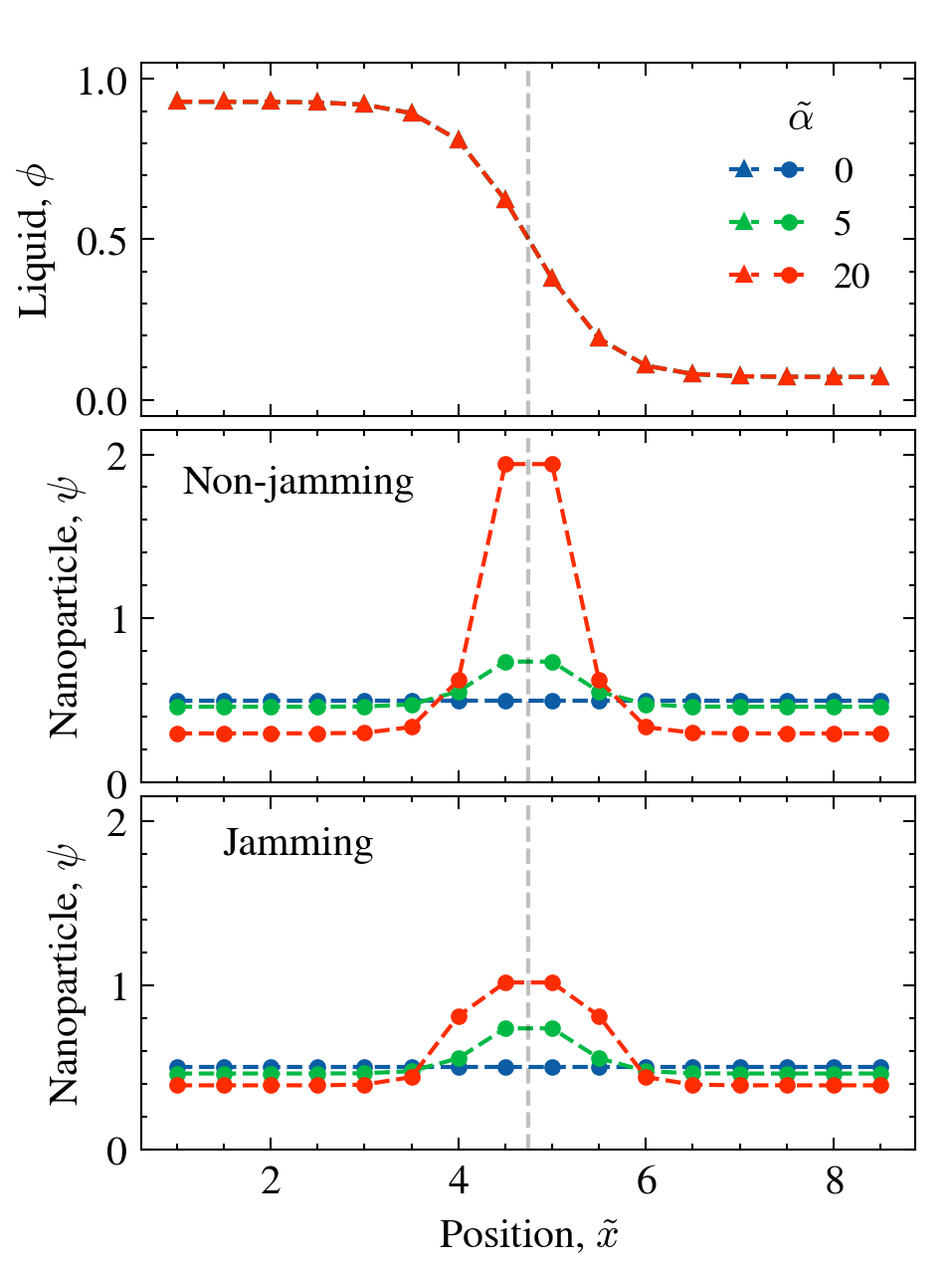}
    \caption{The top and middle panel show the density profiles emerging from the dynamic Eqs.~(\ref{DynamicNoJam1}) and (\ref{DynamicNoJam2}) for a 1D liquid interface with increasing values of the attachment parameter $\tilde{\alpha}$. The liquid ($\phi$) density profiles in the top panel completely overlap, showing no significant distortion. The density profiles for non-jamming and jamming nanoparticles ($\psi$) are shown in the middle and bottom panel, respectively. The latter profiles follow from Eq.~(\ref{DynamicJam2}), where the nanoparticle-dependent mobility $\tilde M_\psi(\psi)$ is calculated \textit{via} the general mobility function (\ref{MobilityFunction}) for $\psi_c=0.95$. The liquid profiles from the coupled Eq.~(\ref{DynamicJam1}) are identical to those shown in the top panel and therefore omitted here.}
    \label{Profiles_JammedAndNonJammed}
\end{figure}
This behaviour is validated by the component density profiles for a 1D liquid interface, shown in the top and middle panel of Figure~\ref{Profiles_JammedAndNonJammed}. Even for the high attachment parameter  $\tilde\alpha=20$ the liquid interface remains undistorted, showing complete overlap with the interface with $\tilde\alpha=0$. At the same time, this high value of $\tilde\alpha$ enables a much more extensive accumulation of nanoparticles, exemplified by the considerably higher value of $\psi$ compared to the previous case in Figure~\ref{CH_Profiles&IFT}A.

With the liquid interface remaining unaffected upon nanoparticle attachment, the next step is to implement the arrest of phase separation. Here, this is achieved by introducing density-dependent mobilities for both the liquid and the nanoparticles, respectively $\tilde M_\phi(\psi)$ and $\tilde M_\psi(\psi)$. In particular, the mobilities sharply drop after the nanoparticle density exceeds a certain threshold value $\psi_c$. This reflects the jamming of a dense layer of interfacial nanoparticles that subsequently pins the liquid to it. The simultaneous nanoparticle jamming and liquid pinning can be mathematically represented through the generalised mobility function 
\begin{equation}
    \tilde M_i(\psi)=\frac{\tilde M_i^0}{2}\left(1-\tanh\left(n\left(\psi-\psi_c\right)\right)\right),
    \label{MobilityFunction}
\end{equation}
where $n$ is a control parameter for the sharpness of the jamming transition, generally taken as $n=75$ for the simulations in this work. In addition, $\tilde M_i^0$ is a constant for the relative base mobility of the component. Scaling the system with respect to the liquid then results in $\tilde M^0_\phi=1$ and $\tilde M_\psi^0= M^0_\psi/M^0_\phi$ for the relative mobilities of the liquid and the nanoparticles. Considering a typical nanoparticle size of 20 nanometers, the latter mobility is subsequently taken as $\tilde M_\psi^0=10^{-2}$, with details provided in the ESI.

The mobility function (\ref{MobilityFunction}) can subsequently be incorporated into Eqs.~(\ref{DynamicNoJam1}) and (\ref{DynamicNoJam2}), resulting in 
\begin{equation}
    \frac{\partial\phi}{\partial \tilde{t}}=\tilde{M}_\phi(\psi)\tilde\nabla^2\tilde{\mu}_\phi+\tilde{\nabla}\tilde{\mu}_{\phi}\cdot\tilde{\nabla}\tilde{M}_\phi(\psi);
    \label{DynamicJam1}
\end{equation}
\begin{equation}
    \frac{\partial\psi}{\partial \tilde{t}}=\tilde{M}_\psi(\psi)\tilde\nabla^2\tilde{\mu}_\psi+\tilde{\nabla}\tilde{\mu}_{\psi}\cdot\tilde{\nabla}\tilde{M}_\psi(\psi),
    \label{DynamicJam2}
\end{equation}
with the chemical potentials of the liquid $\tilde{\mu}_\phi$ and nanoparticles $\tilde{\mu}_\psi$ as
\begin{equation}
    \tilde{\mu}_\phi=\ln{\frac{\phi}{1-\phi}} +\chi(1-2\phi)-\tilde\nabla^2\phi ;
\end{equation}
\begin{equation}
    \tilde{\mu}_\psi=\ln\psi-\frac{\tilde\alpha}{2}\lvert\tilde\nabla\phi\rvert^2.
\end{equation}
Note that in contrast to Eqs.~(\ref{DynamicNoJam1}) and (\ref{DynamicNoJam2}), the density-dependent mobilities of the liquid $\tilde{M}_\phi(\psi)$ and the nanoparticles $\tilde{M}_\psi(\psi)$ in Eqs.~(\ref{DynamicJam1}) and (\ref{DynamicJam2}) are no longer constants across space. Consequently, to maintain the mass conservation inherent to Eqs.~(\ref{CahnHilliard1}) and (\ref{CahnHilliard2}), additional terms are required that account for this spatial dependence.

Through these modifications, the dynamic Eqs. (\ref{DynamicJam1}) and (\ref{DynamicJam2}) capture both the interfacial accumulation and jamming of the nanoparticles, in addition to the established behaviour of the liquids. This is illustrated by the density profiles in the bottom panel of Figure~\ref{Profiles_JammedAndNonJammed}. As the liquid interfaces emerging from Eq.~(\ref{DynamicJam1}) are identical to the ones shown in the top panel of Figure~\ref{Profiles_JammedAndNonJammed}, only the density profiles for the nanoparticles are shown here.

In contrast to the non-jamming particles in the middle panel of Figure~\ref{Profiles_JammedAndNonJammed}, the interfacial density of the jamming nanoparticles in the bottom panel does not exceed $\psi=1.02$ even for $\tilde{\alpha}=20$. This is because the jamming transition is set to take place at a nanoparticle density of $\psi_c=0.95$ here. Because the jamming transition is not instantaneous, the lowered interfacial mobility causes the nanoparticle concentration to stabilise at a level slightly higher than the jamming threshold $\psi_c$. Consequently, the density profile for $\tilde{\alpha}=20$ effectively reflects the formation of a jammed layer of interfacial nanoparticles. 

Interestingly, for $\tilde{\alpha}=20$ the interfacial density of the jamming nanoparticles is broadened relative to that of the non-jamming ones. The reduced interfacial mobility of the jammed layer at the centre of the interface prevents further transport of nanoparticles from adjacent lattice sites, resulting in local accumulation. This mechanism is supported by the identical density profiles of the jamming and non-jamming particles for $\tilde{\alpha}=5$. In that case, the equilibrium density of the interfacial nanoparticles lies below the jamming threshold $\psi_c$, resulting in equivalent accumulation dynamics. Finally, note that the extent of bulk depletion is lower for the jamming nanoparticles compared with the non-jamming ones; despite the broadening of the profile, the jammed interfacial layer can accommodate fewer nanoparticles from the bulk phases because of its lower interfacial density. 

\section{Results}
With the theoretical foundation established, the emergence of particle-stabilised emulsions is demonstrated in simulations. The subsequent coarsening analysis elucidates the mechanism behind their stability, before finally applying the phase-field model to the experimentally relevant case of bijel formation $\textit{via}$ STrIPS.
\subsection{Formation of Particle-Stabilised Emulsions}
Integrating the mobility function into the dynamic equations yields a system in which liquid-liquid phase separation induces the sequential attachment and jamming of interfacially active nanoparticles. The governing equations of the system thus provide a general framework for investigating the formation of particle-stabilised emulsions, irrespective of morphology. The potential of this framework will be demonstrated by examining the two particle-stabilised emulsions with notably different structures: the Pickering emulsion and the bijel. 
\begin{figure*}
    \centering
    \includegraphics[width=\linewidth]{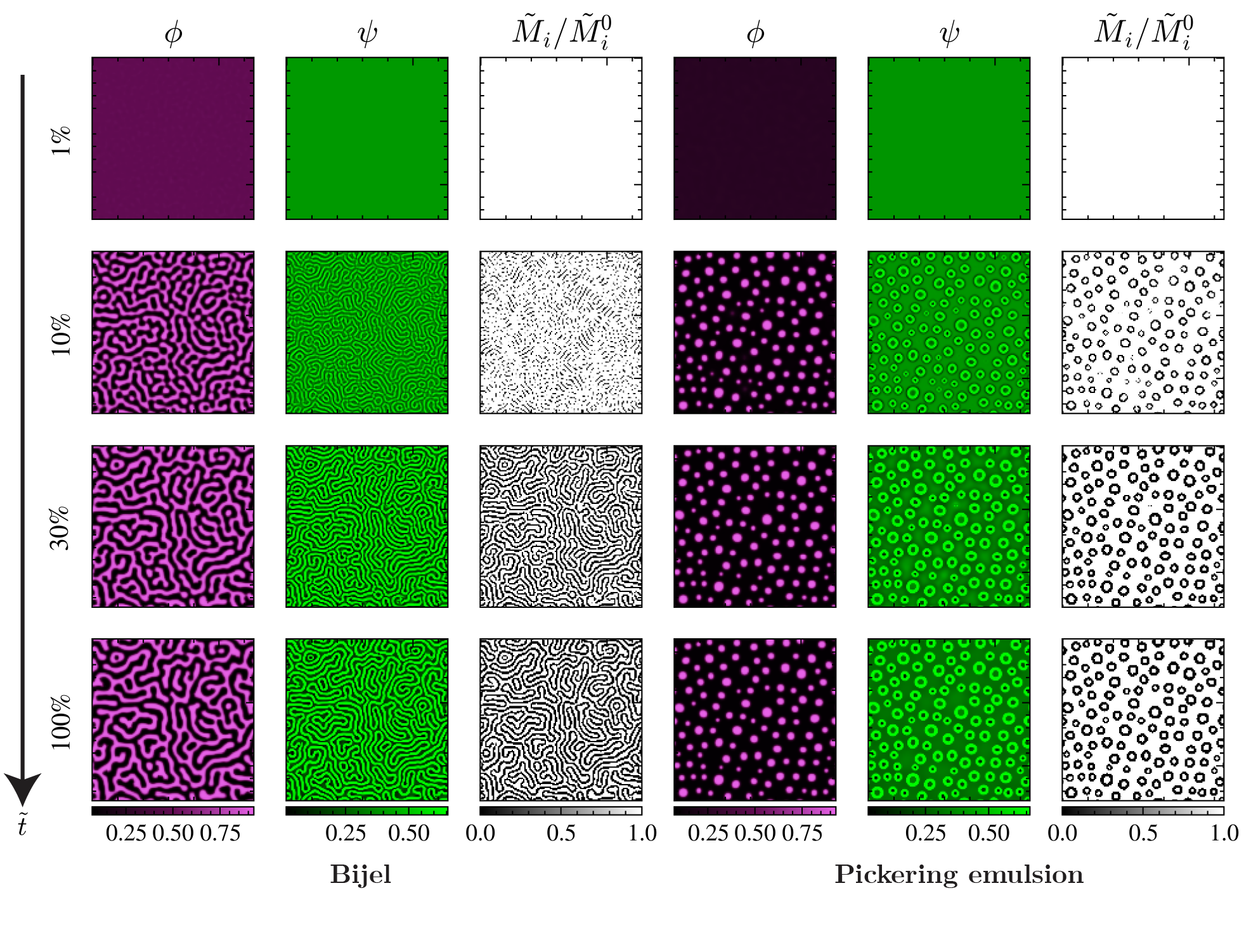}
    \caption{Phase-field simulations showing the formation of the two types of particle-stabilised emulsion: the bijel and the Pickering emulsion. From a homogeneous mixture of immiscible liquids and nanoparticles, with either a critical ($\phi=0.50$) or off-critical ($\phi=0.25$) initial composition of the liquids, phase separation is initiated. The resulting morphologies represent the bijel and the Pickering emulsion, respectively. The state of the liquid $\phi$, nanoparticles $\psi$ and general mobility $\tilde M_i /\tilde M_i^0$ are presented at different stages of the simulation, expressed in percentages of its total duration. This total duration is $\tilde{t}=100$ for the bijel and $\tilde{t}=500$ for the Pickering emulsion, corresponding to their different dynamics of phase separation. Here, the general mobility $\tilde M_i /\tilde M_i^0$ is calculated in accordance with Eq.~(\ref{MobilityFunction}) for $\psi_c=0.60$, while the length of each simulated domain is $\tilde{L}=128$ and the initial concentration of nanoparticles is $\psi_0=0.40$ with $\tilde{\alpha}=60$.}
    \label{EmulsionFormation}
\end{figure*}

The formation dynamics underlying these different particle-stabilised emulsions are illustrated in Figure~\ref{EmulsionFormation}. Starting from a homogeneous mixture of immiscible liquids and nanoparticles, liquid-liquid phase separation is induced through spinodal decomposition. The morphology of the resulting emulsion can be controlled through the initial composition of the liquid mixture. In particular, initiating phase separation from the critical composition $\phi_0=0.50$ gives the labyrinthine network of liquid channels that characterises the bijel structure. The off-critical initial composition $\phi_0=0.25$, however, produces a collection of distinctly separated droplets that effectively represents the structure of a Pickering emulsion. 

The onset of phase separation is accompanied by the attachment of nanoparticles to the newly formed liquid-liquid interface. The nanoparticles then start to accumulate at the interface while gradually depleting from the bulk liquids. The extent of this depletion depends on the morphology, since the interfacial region of the liquid channels is roughly twice that of the dispersed droplets and can support more nanoparticles. Irrespective of morphology, with an initial nanoparticle concentration of $\psi_0=0.40$ the interfacial density quickly exceeds the set threshold value $\psi_c=0.60$. The interfacial mobilities of both the liquids and the nanoparticles are then significantly reduced, corresponding to the jamming of the interfacial nanoparticle layer and the pinning of the liquid interface. The final outcome of this process is a stable structure that does not undergo further coarsening.

\subsection{Stability of Particle-Stabilised Emulsions}
\begin{figure*}
    \centering
    \includegraphics[width=\linewidth]{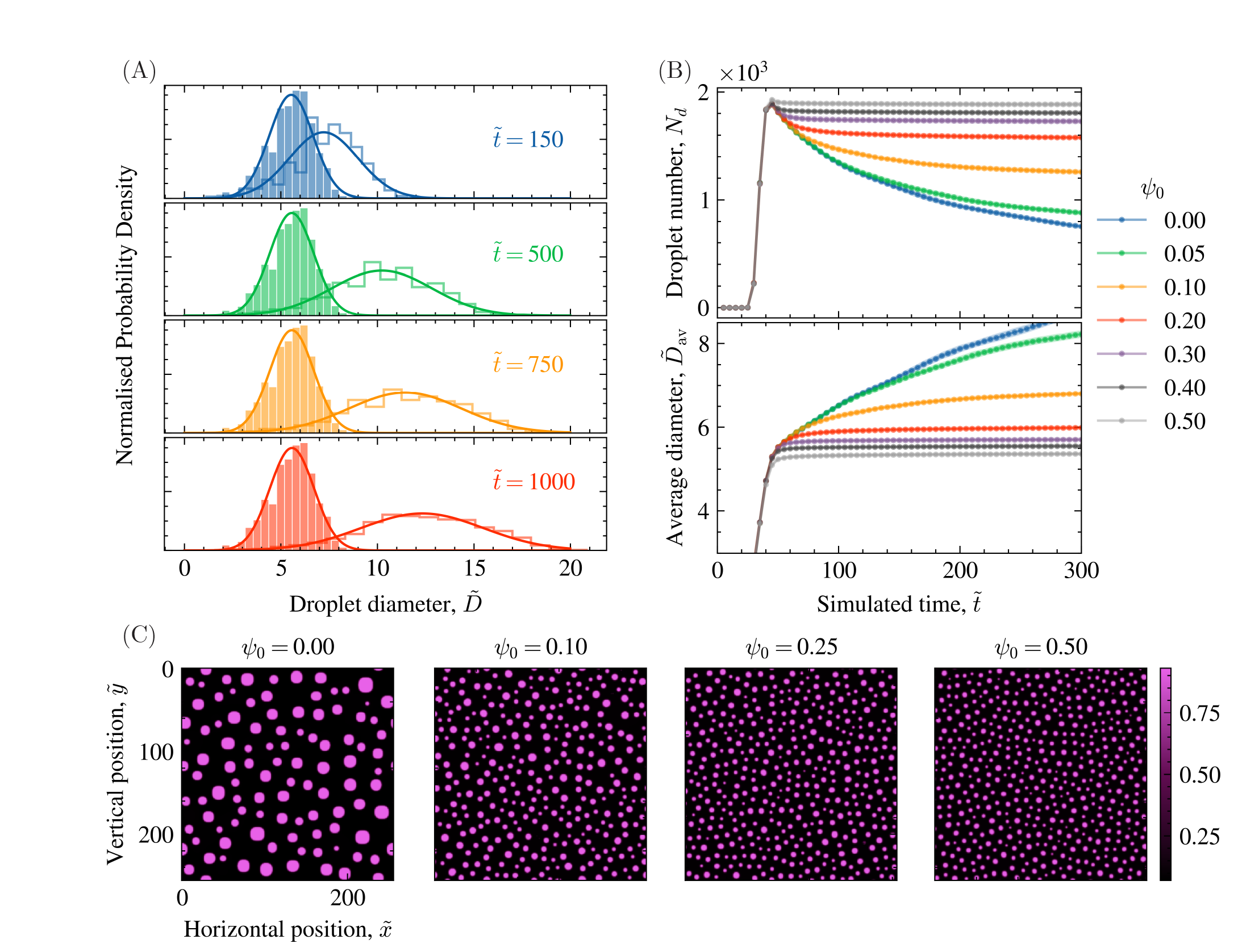}
    \caption{(A) Histograms showing the time-dependent distributions of the droplet diameter $\tilde{D}$ in 2D Pickering emulsions, both with and without nanoparticles. For the Pickering emulsions with nanoparticles, the initial particle density is set to $\psi_0=0.40$ with $\tilde{\alpha}=60$ and $\psi_c=0.60$. In the histograms, filled and empty bars represent the presence and absence of nanoparticles, respectively. While the distributions of Pickering emulsions with nanoparticles show no significant change over time, the absence of nanoparticles results in considerable structural coarsening due to Ostwald ripening. (B) Profiles of the droplet number $N_d$ and the average droplet diameter $\tilde{D}_\mathrm{av}$ during the formation of 2D Pickering emulsions, shown for different values of the initial nanoparticle density $\psi_0$. (C) 2D morphologies representing Pickering emulsions with different initial nanoparticle density $\psi_0$, taken at $\tilde{t}=1000$. Higher nanoparticle densities lead to phase separation being arrested at earlier stages, resulting in Pickering emulsions with smaller, more monodisperse droplets. Note that the simulations in (C) use a system length of $\tilde{L}=256$ for ease of visualisation, while those in (A) and (B) employ a larger value of $\tilde{L}=512$ to improve statistical robustness.}
    \label{DropletCoarsening}
\end{figure*}
The origin of the observed stability for the simulated structures can be readily deduced from their coarsening behaviour. While true for both types of particle-stabilised emulsions, the Pickering emulsion serves as the more illustrative case compared to the bijel. That is, for a Pickering emulsion the characteristic size is simply given by the droplet diameter. Not only is this characteristic size straightforward to determine, it also exhibits a distinct distribution that evolves over time. This is exemplified by the histograms in Figure~\ref{DropletCoarsening}A, showing the distributions of the droplet diameter for simulated Pickering emulsions with and without nanoparticles. 

Without nanoparticles, the distributions in Figure~\ref{DropletCoarsening}A change in both position and shape. Not only do they shift to the right, they also noticeably broaden. This behaviour reflects the coarsening of the Pickering emulsion. Whereas the shift in the distribution indicates a gradual increase in the average droplet size over time, its broadening reveals a simultaneous increase in the polydispersity. These observations all point to Ostwald ripening as the core coarsening mechanism in the system. That is, in this process larger droplets grow at the expense of smaller ones, increasing both the average droplet size and the polydispersity. Ostwald ripening being the dominant mechanism for coarsening is further supported by the fact that these emulsion droplets are well-dispersed and do not undergo Brownian motion.

As is visible in Figure~\ref{DropletCoarsening}A, the distributions for Pickering emulsions with nanoparticles show no significant change over time. Consequently, when the nanoparticles attach to the liquid-liquid interface they effectively arrest any further phase separation and structure coarsening. In this manner, the nanoparticles create a kinetically arrested emulsion in which Ostwald ripening is apparently suppressed. These results are further attested by the profiles in Figure~\ref{DropletCoarsening}B, showing the change in both the number of droplets and their average diameter over time. 

Without nanoparticles, both the number of droplets and their average diameter rapidly increase with the onset of phase separation. After this initial stage of droplet formation, the structure then starts to coarsen, showing features consistent with Ostwald ripening. The dynamics change considerably, however, when nanoparticles are included. While early droplet formation remains unaffected, significant deviations start to occur once nanoparticles attach to the formed liquid interface. In particular, the nanoparticles slow down the change in both droplet number and average diameter, thereby reducing the rate of structure coarsening. This effect becomes more pronounced for higher initial nanoparticle concentration $\psi_0$, eventually causing both the droplet number and the average diameter to reach a constant value. These findings are consistent with the literature on Pickering emulsions, matching the results of both fully particle-resolved simulations\cite{Frijters2014} and experiments\cite{Arditty2003}, yet potential differences in the underlying mechanisms are treated in more detail in the Discussion section. Consequently, the attachment of nanoparticles successfully arrests phase separation in our simulations, protecting the emulsion against further coarsening. 

In addition to preventing Ostwald ripening, the nanoparticles have other pronounced effects on the morphology of the Pickering emulsion. As demonstrated by the profiles in Figure~\ref{DropletCoarsening}B and the 2D structures in Figure~\ref{DropletCoarsening}C, both the average diameter and polydispersity of the particle-stabilised droplets decrease with an increasing initial nanoparticle concentration $\psi_0$. This observation, which again is in qualitative agreement with previous studies on Pickering emulsions\cite{Arditty2003,Frijters2014}, is directly related to the mobility function of the nanoparticles (Eq.~(\ref{MobilityFunction})). Namely, when nanoparticles accumulate on the liquid interface, a higher initial concentration $\psi_0$ simply results in the threshold value $\psi_c$ being reached faster. Jamming of the interfacial nanoparticle layer then also occurs faster, arresting phase separation at an earlier stage. As indicated by the distributions in Figure~\ref{DropletCoarsening}A, this corresponds to smaller and more monodisperse particle-stabilised droplets.

Summarising the findings above, the stability of the simulated particle-stabilised emulsions is a direct consequence of the attachment of nanoparticles to the liquid-liquid interface. In particular, the lowered interfacial mobility associated with the jammed nanoparticles appears to suppress Ostwald ripening, preventing further phase separation and structure coarsening. With the stabilisation mechanism fully elucidated, the model can now be applied to an experimentally relevant system: bijel formation \textit{via} STrIPS. 

\subsection{Bijel Formation \textit{via} STrIPS}
\begin{figure*}
    \centering
    \includegraphics[width=\linewidth]{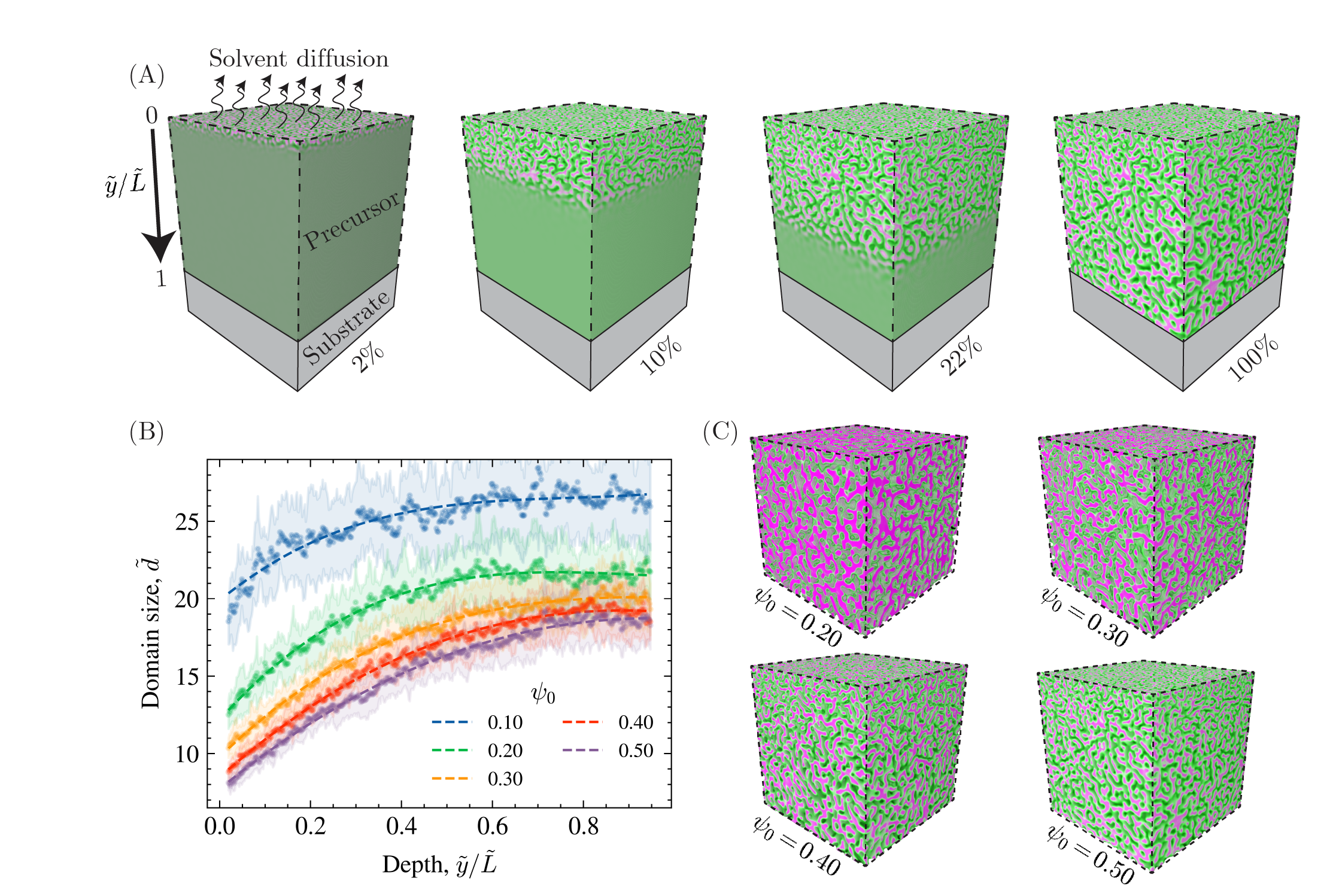}
    \caption{(A) 3D simulation of bijel formation \textit{via} STrIPS. Structures are shown at different percentages of the total simulation time $\tilde{t}=45000$, with more details regarding the simulation parameters provided in the ESI. A precursor mixture containing immiscible liquids, a solvent and nanoparticles is placed between a solid substrate and a liquid ambient phase, respectively located at normalised depths of $\tilde y/\tilde L=1$ and $\tilde y/\tilde L=0$. Here, $\tilde{L}=256$ is the total depth of the simulation. Solvent is only allowed to diffuse into the liquid ambient phase, triggering the spinodal decomposition of the precursor mixture. The nanoparticles subsequently adsorb to the liquid interface, stabilising the structure by preventing further phase separation. The immiscible liquids are shown in black and magenta, while the nanoparticles are presented in green. The solvent is not explicitly depicted. (B) Profiles of the average domain size over the depth of the simulated bijels with $\tilde{L}=512$, shown for different initial concentrations of the nanoparticles $\psi_0$. The shaded regions indicate areas within a single standard deviation from the average, while the dashed lines are third-degree polynomial fits meant to guide the eye. (C) 3D morphologies representing STrIPS bijels with different initial nanoparticle concentrations $\psi_0$. Note that the average domain size decreases with $\psi_0$.}
    \label{STrIPS}
\end{figure*}

Recently, some of us have introduced a phase-field model for STrIPS bijels\cite{Steenhoff2025}. Although this model captures the general dynamics of STrIPS, it remains limited in some aspects. Most notably, it lacks nanoparticles. However, by merging this STrIPS model with the presented phase-field framework for particle-stabilised emulsions, the role of the nanoparticles in arresting phase separation can be properly incorporated. The details of this merger, including the governing equations and simulation parameters, can be found in the ESI. 

In short, an additional field is introduced that represents the solvent and is directly coupled to the interaction parameter $\chi$ between the immiscible liquids. The solvent can exit the system by diffusing across the upper boundary into an ambient liquid phase with zero solvent concentration. Lower solvent levels gradually increase the interaction parameter $\chi$, eventually triggering phase separation of the immiscible liquids. The resulting phase-field model enables investigation into the morphological influence of the nanoparticles on the STrIPS bijels, demonstrated through the results shown in Figure~\ref{STrIPS}. 

Figure~\ref{STrIPS}A shows a 3D simulation of bijel formation \textit{via} STrIPS. A homogeneous precursor mixture, containing immiscible liquids, nanoparticles and a solvent, is placed between a solid substrate and a liquid ambient phase. The solvent then diffuses asymmetrically into the liquid ambient phase, mimicking the conditions of bijel fabrication in roll-to-roll STrIPS\cite{Siegel2024}. The removal of the solvent initiates phase separation through spinodal decomposition, forming the interwoven network of liquid channels that characterises the bijel structure. The nanoparticles subsequently attach to the newly formed liquid interfaces, accumulating there until their density exceeds the jamming threshold. Once jammed, the dense interfacial layer of nanoparticles stabilises the structure by preventing further phase separation. As the solvent is gradually depleted from the precursor mixture, a progressive front of liquid phase separation and nanoparticle stabilisation makes its way through the system. Eventually, this process is completed throughout, resulting in a particle-stabilised emulsion that undergoes no further coarsening. 

The particle-stabilised emulsions produced by these 3D simulations display the hallmark morhological trait of the STrIPS bijel: a pronounced gradient in the domain size that extends throughout the structure. As illustrated by the depth profiles in Figure~\ref{STrIPS}B, the domains are smallest directly adjacent to the ambient liquid and become progressively larger towards the solid substrate. As this gradient is a defining attribute for experimental STrIPS bijels, its presence in our numerical simulations indicates that the phase-field model effectively captures the coupled dynamics of phase separation and nanoparticle attachment. Consequently, the origin of the domain size gradient in the simulated system might provide further insights into the experimental ones.  

In the simulations, the domain size gradient arises because of local differences in the relative rates of solvent-mediated phase separation and nanoparticle attachment. Close to the interface with the ambient liquid, the rapid depletion of solvent causes the formation of distinctly different liquid phases early on. This promotes the fast accumulation of nanoparticles, forming a jammed interfacial layer before significant structural coarsening can take place. Deeper in the precursor, however, solvent levels remain higher for longer. These higher solvent levels slow down the attachment of the nanoparticles while still allowing for the coarsening of the liquid domains. Deeper in the structure, phase separation is thus arrested at a later stage, resulting in larger domains. 

In addition to creating relative differences in the domain size over the structure, the nanoparticles also have a more absolute influence on the bijel morphology. In particular, the profiles in Figure~\ref{STrIPS}B and the 3D structures in Figure~\ref{STrIPS}C indicate that the average domain size across the entire bijel decreases with higher initial concentrations of the nanoparticles $\psi_0$. Here, higher concentrations of nanoparticles induce interfacial jamming at an earlier stage of phase separation, resulting in the stabilisation of smaller domains. This inverse relation between domain size and nanoparticle concentration matches previous findings for bijels\cite{Herzig2007,Jansen2011}, although specific experimental evidence for STrIPS bijels remains lacking. 

Finally, the simulated STrIPS bijels exhibit one other morphological feature that warrants discussion. That is, the slopes of the domain size profiles in Figure~\ref{STrIPS}B mirror those of the solvent concentration during STrIPS: a steep slope directly adjacent to the ambient liquid and a flatter slope deeper in. Moreover, these slopes remain relatively unaffected over a wide range of nanoparticle concentrations, with only the profile at the lowest value of $\psi_0=0.10$ showing a significant deviation. The different profiles thus mostly retain their shape with decreasing nanoparticle concentration, yet still shift upwards to reflect the longer coarsening time prior to structural arrest. For a more detailed analysis relating the gradients in the bijel domain size to the solvent profiles, the reader is referred to the ESI. 

In summary, the phase-field framework of particle-stabilised emulsions was successfully incorporated into the STrIPS model. With this combination, the effect of the nanoparticles on the morphology of the STrIPS bijel was investigated. Simulations reproduced the characteristic size gradient of the STrIPS bijel, while establishing an inverse relation between the overall domain size and the nanoparticle concentration. These results demonstrate the potential of the phase-field model for elucidating the morphological features of complex particle-stabilised emulsions. 

\section{Discussion}

Although the previous section highlights the effectiveness of the phase-field model in dealing with particle-stabilised emulsions, some challenges remain. First, there is the detachment of the dynamic equations from the free-energy functional of the system. While the former are inspired by the form of the latter, they cannot be directly derived from it. This ``non-integrability'' is common practice in fields such as active matter\cite{Wittkowski2014}, yet the formation of a particle-stabilised emulsion ideally lends itself to a more thermodynamically rigorous treatment. The latter treatment would additionally allow for a more extensive description of the liquid-nanoparticle interaction, albeit at a higher computational cost. 

However, the main issue with the thermodynamic approach is the insufficient separation of scales for the nanoparticles and the molecular liquids. That is, including their respective contributions in the full free-energy functional inherently causes them to operate in the same regime. The squared gradient term $|\nabla\phi|^2$ then always imposes a surfactant-like character on the nanoparticles, influencing the interfacial tension of the liquids. Initially, this work attempted to mediate some of the resulting effects. Examples include a higher-order gradient term\cite{Teubner1987} in the free-energy functional, to deal with negative interfacial tensions, or even a gradient-free representation of the liquid interface\cite{Engblom2013}. However, these efforts proved unsuccessful so far. Accordingly, particle-resolved models currently remain the better option for an approach fully predicated on thermodynamics. 

Second, the phase-field model employs some simplifications that warrant caution when comparing its findings with experimental data. Most importantly, it does not consider hydrodynamics. This was initially done for the sake of computational efficiency, given that early phase separation tends to be dominated by diffusion. At later stages of structural coarsening, however, hydrodynamics become increasingly relevant. Furthermore, the inclusion of hydrodynamics is of particular interest for the STrIPS bijel, where the morphological effects of convective flows are poorly understood. While the diffusion-based STrIPS models capture the general trends observed in experiments, analysis of similar systems\cite{Tree2018} indicates that advection likely plays a prominent role as well. Consequently, extending the phase-field model with hydrodynamics provides valuable insight into the experimentally observed structures of STrIPS bijels. 

Similarly, the structural analysis in Figure~\ref{DropletCoarsening} does not capture the full coarsening behaviour displayed by Pickering emulsions. That is, the observed stability is attributed solely to suppressed Ostwald ripening, achieved by the reduced interfacial mobility of the nanoparticle layer. Effectively, this corresponds to completely inhibited diffusion of the liquids across the particle-laden interface. In experimental Pickering-emulsions, however, diffusion can still occur through the interstitial spaces in the particle layer\cite{Liu2024}, facilitating Ostwald ripening even at higher coverage in some cases\cite{Thompson2018,Kumar2023}. 
 
In addition, Pickering emulsions can coarsen through the coalescence of their constituent droplets, particularly at low nanoparticle coverage. Generally, this coalescence is driven by either hydrodynamics or Brownian motion exhibited by the droplets. Since neither of these two mechanisms are included in the current phase-field model, coalescence does not occur in the simulations. An improved version of the phase-field model, reflecting the structural coarsening of an actual Pickering emulsion, should thus include droplet coalescence in some manner. This could be achieved by mimicking the effect of Brownian motion, although the incorporation of thermal fluctuations in the composition should already suffice. Further research is required to verify the latter claim, however. 

Next, it is worth critically evaluating the apparent agreement between phase-field simulations and experiments. Specifically, whether the similar trends in simulation and experiment actually share a common origin. For example, consider the inverse relationship between the domain size and the nanoparticle concentration that is found for the bijel. Experiments indicate that this inverse relationship results from the limited availability of nanoparticles at relatively low concentrations\cite{Herzig2007}. That is, higher nanoparticle concentrations can stabilise a larger interfacial area, corresponding to smaller domains. In contrast, the simulations attribute the smaller domains at higher nanoparticle concentrations to a faster jamming transition, arresting phase separation at an earlier stage. Consequently, in simulations the origin of this trend appears to be kinetic in nature, rather than being related to nanoparticle availability. 

However, it must be noted that these simulations are performed with comparatively high nanoparticle concentrations, reflecting the experimental conditions of the STrIPS bijel. At these high concentrations, the kinetics of nanoparticle attachment could also become relevant for the experimental systems, but supporting this claim requires more experimental evidence. Conversely, at low concentrations the availability of nanoparticles might be the determining factor for the domain size in simulations. While this is not readily apparent from the analysis in Figure~\ref{STrIPS}, it can be verified as follows. First, a measure of the interfacial area needs to be calculated during liquid phase separation. Then, it has to be determined whether the available nanoparticles are sufficient to stabilise this interfacial area when given enough time. However, such an analysis lies outside the scope of this work. 

Finally, the phase-field model would benefit from further refinement to open up new avenues for research. For example, the free-energy description of the nanoparticles can be expanded to encode properties such as their size, shape, and contact angle, at least on an effective level. In this manner, the phase-field model facilitates the investigation of these properties for particle-stabilised emulsions on larger scales. Alternatively, wetting effects can be included for the phase-separating liquids. This is particularly relevant for STrIPS bijels, whose morphologies exhibit a marked dependence on the wetting of solid substrates in experiments\cite{Siegel2025}. Regardless, both options highlight the versatility of the phase-field model in dealing with the various aspects involved in the formation of particle-stabilised emulsions. 

\section{Conclusions}
In this work, a phase-field model for particle-stabilised emulsions is introduced. Based on the thermodynamics of an immiscible liquid-nanoparticle mixture, differential equations are derived that describe the coupled dynamics of liquid phase separation and nanoparticle adsorption. These initial equations, however, result in severe, unphysical deformation of the liquid interface upon nanoparticle accumulation. To address this issue, a revised set of equations is presented that partially decouples the dynamics of the liquids and the nanoparticles. These dynamic equations capture the same overall behaviour without suffering from interfacial distortions. 

Subsequently, it is demonstrated that this set of governing equations results in the formation of a particle-stabilised emulsion. Reflecting the jamming of interfacial nanoparticles, phase separation is arrested by locally modulating the interfacial mobility upon nanoparticle accumulation. Coarsening analysis then reveals that the imposed stability is caused by an effective suppression of Ostwald ripening. 

Finally, the phase-field model is applied to investigate bijel formation \textit{via} STrIPS. Here, it is shown that the coupled dynamics of solvent-mediated phase separation and nanoparticle attachment produce a distinct gradient in the domain size throughout the bijel. In addition, it is found that the nanoparticle concentration significantly influences the bijel morphology, with higher concentrations being associated with smaller average domains. 

Together, our findings illustrate the effectiveness of the presented phase-field framework for particle-stabilised emulsions. By resolving the interplay between liquid phase separation and nanoparticle attachment on a large scale, the phase-field model opens novel avenues for research into the morphological features of particle-stabilised emulsions. 

\section*{Data Availability Statement}
The \texttt{Python} scripts and datasets that support the findings of this study are openly available on \href{https://github.com/JesseSteenhoff/PhaseFieldModelsForParticleStabilisedEmulsions}{Github} and archived on Zenodo at https://doi.org/10.5281/zenodo.17668890. 

\begin{acknowledgments}
This publication is part of the project ‘‘Bijel templated membranes for molecular separations'' (with project number 18632 of the research programme Vidi 2019), which is financed by the Dutch Research Council (NWO). Joost de Graaf acknowledges funding through NWO grant OCENW.KLEIN.354. 
\end{acknowledgments}

\section*{Author Contributions}
\textbf{Elisabeth C. Eij:} Formal Analysis; Methodology; Investigation; Software; Writing- Review \& Editing. \textbf{Joost de Graaf:} Conceptualisation;  Methodology; Supervision; Writing- Review \& Editing. \textbf{Martin F. Haase:} Conceptualisation; Funding Acquisition; Supervision; Writing- Review \& Editing. \textbf{Jesse M. Steenhoff:} Conceptualisation; Formal Analysis, Investigation; Methodology; Supervision; Software; Validation; Visualisation; Writing - Original Draft; Writing- Review \& Editing. 

\bibliography{MainText}

\end{document}